\documentstyle[aps,prl,twocolumn,epsfig]{revtex}

\def\comment#1{}

\def\ps#1{\begin{center}\leavevmode\hbox{\epsfxsize=2.7in\epsfysize=3.3in\epsfbox{#1}}\end{center}}

\newcommand{\be}{\begin{equation}}
\newcommand{\ee}{\end{equation}}
\def\bea {\begin{eqnarray}}
\def\eea {\end{eqnarray}}

\def\ell{i}

\begin{document}
\title{Microscopic study of inhomogeneous superconductors}
\author{K. Tanaka$^{1,2}$ and F. Marsiglio$^3$}
\address{$^1$Department of Physics and Engineering Physics, 
University of Saskatchewan,\\ Saskatoon, SK, Canada S7N 5E2}
\address{$^2$Materials Science Division, Argonne National Laboratory,\\
9700 South Cass Avenue, Argonne, IL 60439}
\address{$^3$Department of Physics, University of Alberta,
Edmonton, Alberta, Canada T6G 2J1}
\date{published in Journal of Physics and Chemistry of Solids {\bf 63} (2002)
  2287}
%\twocolumn[\hsize\textwidth\columnwidth\hsize\csname@twocolumnfalse\endcsname
\maketitle
\begin{abstract}
We study various inhomogeneity effects on superconductivity as due to
quantum confinement, surfaces and impurities, using the self-consistent
Bogoliubov-de Gennes formalism on the attractive Hubbard model.
The results are also compared with those obtained from the Anderson
prescription, a BCS formalism for incorporating spatial
inhomogeneity.
\end{abstract}
%]
\narrowtext

Recent improvements in Scanning Tunneling Microscopy (STM) allow unprecedented
probing of the surface properties of metals and superconductors \cite{hudson01}.
On the other hand, theoretical work, beginning with BCS \cite{bardeen57}, has,
in the past, most often utilized the bulk approximation. This allows the problem to
be formulated in momentum space, where considerable simplification can be gained.
More recently, work has focused more on spatially inhomogeneous problems, and
the quasiclassical \cite{rainer96} or Bogoliubov-de Gennes (BdG) 
\cite{degennes66,flatte99} equations have been utilized.

In this study we examine some simple manifestations of translational invariance
breaking, namely surfaces and point impurities. This is done in the context of
s-wave superconductivity. Following a brief description of our modelling and 
formulation, we present some results in one and two dimensions.
We find that the surface characteristics of superconductivity often differ
considerably from those in the bulk.

To model a finite system we use a tight-binding formulation. 
Here the kinetic energy is parameterized by a hopping matrix element ($t$), 
which allows an electron to move from one ion to the next. 
The surface is modelled, in this framework, in terms of
open boundary conditions (OBC) where $t=0$ from the surface site to outside,
as opposed to periodic boundary conditions (PBC), as for modelling 
homogeneous systems.  In reality
one can attempt a much more sophisticated description, since, for example, one
would expect a modification of the electronic orbitals themselves near a 
surface (or impurity). Then the hopping parameter and the other interaction 
parameters ought to be modified. We omit these finer possibilities.

Impurities are modelled simply by energy level changes at the impurity site. In
this work we always take the impurity site to be one of the lattice sites. 
These are meant to model `normal' impurity scattering, i.e., with no spin flip.

The s-wave superconductivity is most simply described by the attractive 
Hubbard model, with uniform attractive interaction $|U|$. 
The BdG equations are readily formulated
for this problem \cite{tanaka00}, and yield a site-dependent order parameter, 
$\Delta_i$, a site-dependent electron density, $n_i$, as well as a frequency 
and site-dependent spectral function, $A_i(\omega)$. 
This latter quantity can be measured directly by STM.
We have also formulated BCS-like equations \cite{tanaka00} which we
call the `Anderson prescription' \cite{anderson59}. 
Results from this calculation will also be shown.

Near a surface or an impurity, both the order parameter and the density
distribution exhibit ``Friedel-like'' oscillations.
In Fig.~1 this is demonstrated for the case of a surface, for 
a one-dimensional 128-site system with coupling strength
$|U|/t=1.2$ for various values
of the average density $n$.  The order parameter $\Delta_i$
is shown as a function of site number $i$ for half the system size,
where site 1 is a surface.
The BdG results (solid curves) are compared with the Anderson results
(dashed curves).  
A surface has an effect of pair breaking in the sense that the order
parameter is forced to be zero right outside the surface.  
Thus the order parameter shows ``Friedel-like'' oscillations near the
surface, roughly over the coherence length scale.
These oscillations reflect the single-particle wave function at
the Fermi level, and for a given density $n$,
their period in site number is governed by 
$\pi / k_F a$, where $a$ is the lattice constant.
This is seen clearly in Fig.~1.  Note also that the length scale 
over which the oscillations decay also decreases as $n$ decreases:
this is due to the decreasing coherence length.

In Fig.~2 the order parameter $\Delta_i$ and the density distribution $n_i$
are shown for an $N=64$ chain with PBC and an impurity potential
$\epsilon_{32}=-0.5 t$ in the middle, for $n=0.9$.
When the impurity potential is attractive and relatively weak (e.g., 10\% of
the kinetic energy), 
the gap parameter is enhanced at the impurity site \cite{tanaka00}.  
For a repulsive or stronger attractive potential (e.g., the case shown
in Fig.~2) the gap is suppressed at that site, while the density is
peaked there.  In either case, both $\Delta_i$ and $n_i$ show the 
``Friedel-like'' oscillations around the impurity site.
We indicate in Fig.~2 the coherence length evaluated by the BCS expression,
$\xi=\sqrt{\langle R^2\rangle}$,
where $\langle R^2\rangle$ is the mean square radius of an electron pair 
\cite{tinkham75,marsiglio90}, for each coupling strength.
At low temperatures (as is the case here), this $\xi$ coincides with the
Ginzburg-Landau coherence length and determines the length scale
of order parameter fluctuations.
In fact, for $|U|/t=1.1$, $\xi$ is about half the system size for $N=64$
and we have interference effects due to finite size: for $|U|/t=2$,
both the gap and density distribution are seen to relax to their bulk
value far away from the impurity.

In Fig.~2 again the BdG and Anderson results are plotted in solid and dashed
curves, respectively.
As can be seen in both Figs.~1 and 2, the Anderson prescription captures
the essential features of the ``Friedel-like'' oscillations and reproduces the
BdG results remarkably well.  
An exception is the case of intermediate to strong coupling 
with a weak impurity potential \cite{tanaka00}, where the Anderson
approach tends to give the order parameter 
almost independent of position and underestimate the oscillation amplitudes.
This is apparent in the lower two panels of Fig.~2 for $|U|/t=2$:
it can also be seen that the density peak at the impurity site 
is underestimated.

We further compare the BdG and Anderson results at an impurity site
in Fig.~3, for a 32-site chain with PBC and $|U|/t=1$ for $n=0.8$.
Here the order parameter and the density are plotted as a function
of the impurity potential at site 16.  The calculated points
(crosses and stars for the BdG and Anderson results, respectively)
are connected by interpolation.
In the weak-coupling limit (as is the case here), 
the Anderson prescription gives the overall 
behaviour of the order parameter and density distribution 
as a function of position correctly.
For relatively strong impurity potential, however, the Anderson results
tend to deviate from the BdG ones at the impurity site.
As the impurity potential becomes extremely strong, 
the two results at that site tend to agree, regardless of the coupling
strength.  

The Anderson approach can also capture detailed changes 
in the local density of states (LDOS) around a surface or an impurity.
This is illustrated in Fig.~4 for a one-dimensional 128-site system with OBC
and $|U|/t=2$ at quarter filling, where
the LDOS (with a Gaussian smoothing width $0.05 t$) is shown for various sites
including the surface (site 1).
As can be seen in this figure,
the LDOS at and near a surface is quite different from that in the
bulk (e.g., the one at site 64 or the average DOS in Fig.~4).
In particular, the BCS coherence peaks are not very prominent, 
or almost absent: compare the LDOS for site 1 to 4 with the bulk DOS
in Fig.~4.  
Nonetheless the energy gap in the spectrum on the surface is almost
the same as the bulk value (an exception is in the dilute limit, where
the electron density is greatly reduced near the surface, so the 
spectrum differs considerably from that of the bulk as well \cite{tanaka01}).
On the other hand, it is interesting to note that for quarter filling,
the energy gap is larger at every fourth site (site 4 in Fig.~4).
The Anderson prescription reproduces these features very well, 
while it tends to overestimate the coherence peaks at and near a surface.

Next we illustrate the ``Friedel-like'' oscillations in the order parameter 
in two-dimensional systems.
Due to the local nature of the Hubbard interaction, 
most of the basic features seen in the one-dimensional case apply in higher
dimensions.  
In Fig.~5 the BdG results for the order parameter are shown
for an $N=32\times 32$ system with OBC at half filling,
for $|U|/t=4$ and 1.5.  
Here the peripherals are surfaces, and 
in the ``Friedel-like'' oscillations arising from the surfaces, 
we see the interference of degenerate single-particle states 
at the Fermi level.
The resulting structure of the order parameter can be quite complicated;
however, it is relatively simple at
half filling due to the particle-hole symmetry.
For half filling, with strong coupling (e.g., $|U|/t=4$ in Fig.~5)
the order parameter is constant inside and
larger at and near surfaces.
With weak coupling,
it exhibits  ``Friedel-like'' oscillations along the diagonals, which
become more pronounced and extend over longer ranges as coupling is reduced.
For small systems we then encounter finite size effects.
This is almost the case for $|U|/t=1.5$ in Fig.~5,
and further demonstrated in Fig.~6 for $N=30\times 20$, for 
for $|U|/t=1.2$ and 1.

Using a simple model for s-wave superconductivity, we have used the
self-consistent BdG and Anderson prescriptions to calculate the 
spatially inhomogeneous 
order parameter and local density of states (LDOS), as a function of
average electron density, coupling strength, and impurity potential
strength. We find a significant modification of the order parameter
near a surface or impurity, and over a large parameter regime, the
Anderson prescription works as well as the BdG formulation. In two
dimensions the order parameter oscillates along the diagonals at
half filling and in the weak coupling limit.

The oscillations are a feature captured by both the BdG and Anderson 
prescriptions; they signify another length scale (other than the
coherence length) over which the order parameter can vary. Such a
length scale can be present (and significantly smaller than the
coherence length) because the BdG equations have both pairing
aspects and single-electron aspects within them. Because of the
existence of a Fermi surface the latter lead to the oscillations
we have found. This shares a common origin with the Friedel oscillations
that occur around a single impurity. For this reason we referred to 
these as `Friedel-like' oscillations.

This research was supported by the
Natural Sciences and Engineering Research Council of Canada and
the Canadian Institute for Advanced Research.

\begin{figure}
\ps{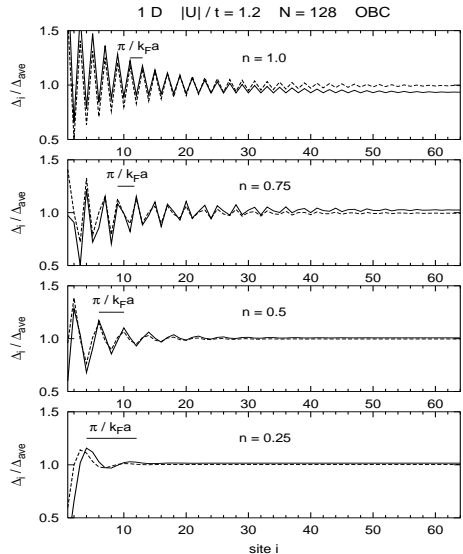}
\caption{Order parameter $\Delta_i$ normalized by the average value 
as a function of site number $i$, for a $128$-site chain with OBC 
with $|U|=1.2 t$, for various average electron densities $n$.
The BdG and Anderson results are plotted with solid and dashed curves, 
respectively.  The period of the ``Friedel-like'' oscillations
in site number increases as $n$ decreases, and 
is consistent with $\pi / k_F a$.}
\end{figure}

\begin{figure}
\ps{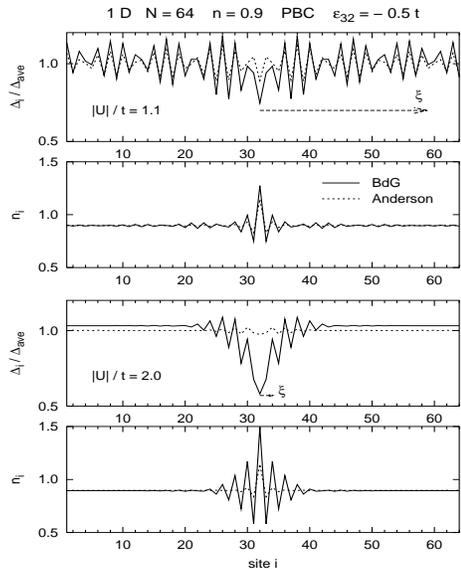}
\caption{
The order parameter $\Delta_i$ and the density distribution $n_i$
are shown for a 64-site chain with PBC and $\epsilon_{32}=-0.5t$ for $n=0.9$.
The upper two panels and the lower ones are for $|U|/t=1.1$ and 2,
respectively.  The coherence length roughly determines the length scale 
over which the ``Friedel-like'' oscillations decay.
The Anderson approach (dashed curves) fails to reproduce the BdG results 
(solid curves) in the strong-coupling limit.
}
\end{figure}

\begin{figure}
\ps{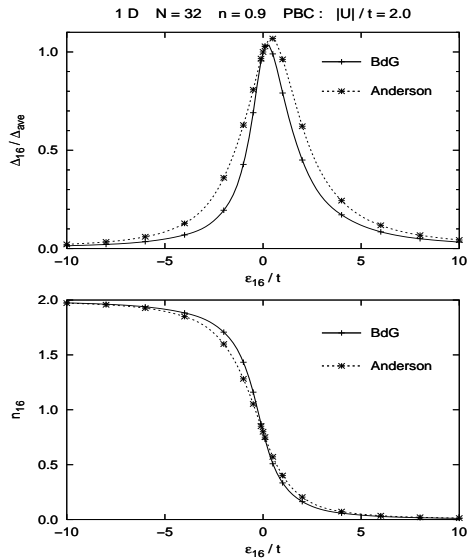}
\caption{
The BdG and Anderson results are compared for the order parameter and
the density at the impurity site as a function of the impurity potential,
$\epsilon_{16}$, for a 32-site chain with PBC and $|U|/t=1$ for $n=0.8$.
}
\end{figure}

\begin{figure}
\ps{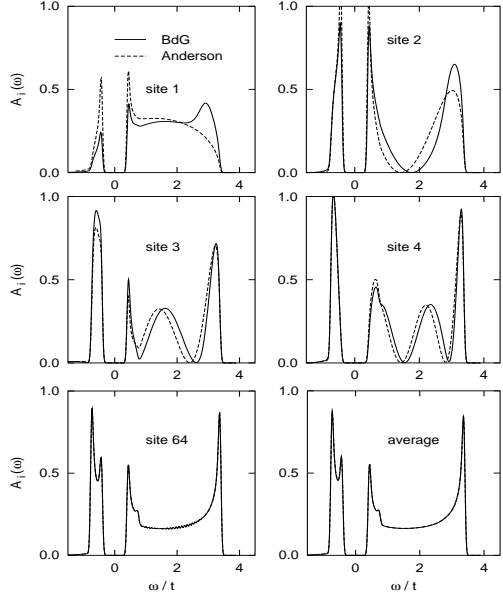}
\caption{
The LDOS for a $128$-site chain with OBC and $|U|/t=2.0$ at quarter filling,
for several sites: site 1 is a surface and site 64 is 
the middle of the sample. The Anderson prescription (dashed curves) 
tends to overestimate the coherence peaks of 
the the BdG results (solid curves).
Near a surface, the LDOS is quite different from that in the bulk 
(site 64 or average), and at quarter filling, 
the energy gap is larger at every fourth site (e.g., site 4).
}
\end{figure}

\begin{figure}
\ps{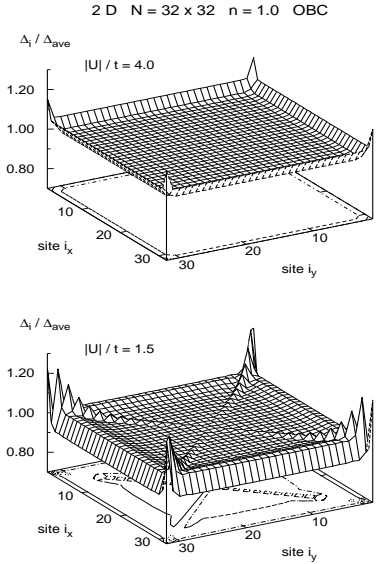}
\caption{
The BdG results for the order parameter for an $N=32\times 32$ system 
with OBC at half filling are shown 
for $|U|/t=4$ and 1.5.  
In the strong-coupling limit, the order parameter is flat inside the sample
but larger at and near surfaces.  For weak coupling it tends to be smaller
at surfaces, while it exhibits the ``Friedel-like'' oscillations:
the latter exist only along the diagonal due to interference effects.
}
\end{figure}

\begin{figure}
\ps{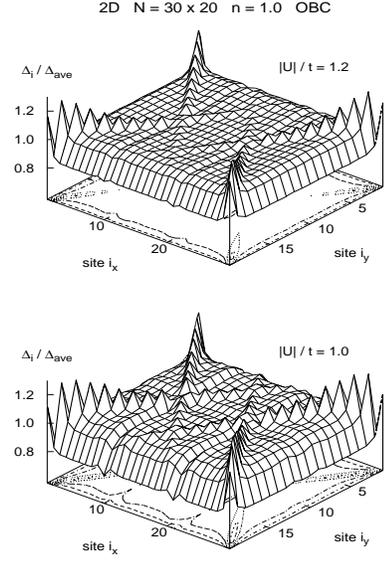}
\caption{
Same as in Fig.~5, but for $N=30\times 20$ and for $|U|/t=1.2$ and 1.  
The ``Friedel-like'' oscillations can be seen more prominently.
}
\end{figure}


\begin{thebibliography} {999}

\bibitem{hudson01} E.W. Hudson, K.M. Lang, V Madhavan, S.H. Pan, 
H. Eisaki, S. Uchida and J.C. Davis, Nature {\bf 411}, 920 (2001).

\bibitem{bardeen57}
J. Bardeen, L.N. Cooper and J.R. Schrieffer, Phys. Rev. {\bf 106}, 162 (1957);
Phys. Rev. {\bf 108}, 1175 (1957). 

\bibitem{rainer96} {\it Quasiclassical Methods in Superconductivity
and Superfluidity}, edited by D. Rainer and J. A. Sauls (Verditz lecture
notes, 1996), unpublished. 

\bibitem{degennes66} P. G. de Gennes, {\it Superconductivity of
Metals and Alloys} (W.A. Benjamin, Inc. New York, 1966).

\bibitem{flatte99} M.E. Flatt\'e and J.M. Byers, in
{\it Solid State Physics} {\bf 52}, 137 (1999). 

\bibitem{tanaka00}
K. Tanaka and F. Marsiglio, Phys. Rev. B{\bf 62} 5345 (2000).

\bibitem{anderson59} P.W. Anderson, J. Phys. Chem. Solids {\bf 11} 26 (1959).

\bibitem{tinkham75} M. Tinkham, {\it Introduction to Superconductivity},
(McGraw-Hill, New York, 1975).

\bibitem{marsiglio90} F. Marsiglio and J.E. Hirsch, Phys. Rev. B{\bf 41}
6435 (1990). 

\bibitem{tanaka01} K. Tanaka and F. Marsiglio, unpublished.

\end{thebibliography}
\end{document}